\documentclass[twocolumn,twoside,slac_two]{revtex4}
\usepackage{graphicx}
\usepackage{fancyhdr}
\begin{document}

\title{Probing proton acceleration in W51C with MAGIC}

%

\author{J. Krause}
\affiliation{Max-Planck-Institut f\"ur Physik, D-80805 M\"unchen, Germany}
\email{julkrau@googlemail.com}

\author{E. Carmona}
\affiliation{Centro de Investigaciones Energ\'eticas, Medioambientales y Tecnol\'ogicas (CIEMAT), Madrid, Spain}

\author{I. Reichardt}
\affiliation{IFAE, Edifici Cn., Campus UAB, E-08193 Bellaterra, Spain}

\author{on behalf of the MAGIC Collaboration}

\begin{abstract}
Located in a dense complex environment, W51C provides an excellent scenario to probe
accelerated protons in SNRs and their interaction with surrounding target material.
Here we report the observation of extended Very High Energy (VHE) gamma-ray emission
from the W51C supernova remnant (SNR) with MAGIC. Detections of extended gamma-ray
emission in the same region have already been reported by the Fermi and H.E.S.S.
collaborations. {\em Fermi}/LAT measured the source spectrum in the energy range between 0.2
and 50 GeV, which was found to be well fit by a hadronic neutral-pion decay model.
The VHE observations presented here, obtained with the improved MAGIC stereo system,
allow us to pinpoint the VHE gamma-ray emission in the dense shocked molecular cloud
surrounding the remnant shell. The MAGIC data also allow us to measure, for the first
time, the VHE emission spectrum of W51C from the highest {\em Fermi}/LAT energies up to
TeV. The spatial distribution and spectral properties of the VHE emission suggest a
hadronic origin of the observed gamma rays. Therefore W51C is a prime candidate for a
cosmic ray accelerator.
\end{abstract}

\maketitle

\thispagestyle{fancy}



\section{Introduction}
Supernova Remnants
(SNRs) are believed to be the sources of galactic cosmic rays for many years.
The underlying process from a theoretical point of view is the diffusive shock
acceleration in which charged particles are accelerated in the expanding shocks of the SNRs, receiving part of the
kinetic energy of the shock through the first order Fermi acceleration~\cite{Reynolds2008}. 
Even though the last years provided increasing support by observational evidences,
there is still no absolute proof that SNRs can accelerate hadronic particles.
The observations of high
energy (HE, $>100\,\mathrm{MeV}$) and very high energy (VHE, $>100\,\mathrm{GeV}$) gamma rays from many SNRs clearly 
showed that they are sites of relativistic particles of energies compatible to the emission seen 
in gamma rays. To unambiguously establish the acceleration of hadronic particles, the
hadronic origin of the observed gamma rays needs to be proven. 
In almost all cases, electromagnetic scenarios (gamma rays produced by
up-scattering of photons through Inverse Compton scattering by
accelerated electrons) and
hadronic scenarios (gamma rays produced after the decay of $\pi^0$)
cannot be distinguished by looking only at the VHE gamma-ray
regime. Therefore observations at other wavelengths are needed to better
understand the production mechanism of gamma rays. This includes on the one hand measurements of the parameters
describing the interstellar medium (ISM) around the object like magnetic field and density.
On the other hand observations of HE gamma rays in the range from MeVs to GeVs are very
important because they can reveal features in the spectrum
that can distinguish both possibilities.
Since the luminosity of gamma ray emission of a hadronic origin is proportional to the product of CR density and the density
of the surrounding ISM, the existence of dense molecular clouds close to the SNR acts as a natural amplifier for the hadronic component.  
A tool to detect evidence of cosmic rays acceleration is therefore provided by gamma rays emitted
by molecular clouds adjacent to or shocked by SNRs~\cite{Hinton2010}. 
W28~\cite{HESS-W28} and IC443~\cite{MAGIC-IC443} are two examples
where this effect might be responsible for the production of VHE
gamma rays. In the W28 case, gamma rays might be produced by cosmic
rays that escaped the accelerating SNR while in IC 443
a {\em Crushed Cloud} scenario might be at work where the molecular cloud is shocked by the
blast wave region of the SNR~\cite{Uchiyama2010}. 
A major differnce in the two scenarios is the fact that the illumination of molecular clouds by runaway cosmic rays
is affected by the energy dependent escape and diffusion of cosmic rays and might therefore show spectral features due to this.
In the case of a {\em Crushed Cloud} the diffusion effects are thought to be much smaller and the emission could be produced by the cosmic ray source spectrum
itself. On the other hand re-acceleration of cosmic rays in the shocked cloud may play a role in this case. 

The SNR W51C is a promising candidate to test and study cosmic ray acceleration in SNRs. 
W51C is a mixed morphology
SNR with an elliptical shape in radio and a size of $0.8^\circ \times
0.6^\circ$.  It is located in the tangential point of the Sagittarius
arm at a distance of $\sim6$~kpc~\cite{Koo1995} and the estimated age
is around 30~kyrs. 
The W51 complex consists of three main components: Two star-forming
regions, W51A and W51B, and the SNR W51C. While W51A is separated from
the other two, W51B overlaps with the North-Western rim of W51C. 
Shocked atomic and molecular gases have
been observed in radio data~\cite{Koo1997a}~\cite{Koo1997b}, providing
direct evidence on the interaction of the W51C shock with a large
molecular cloud. This strongly indicates that W51C represents a {\em Crushed Cloud} scenario rather than a cloud illumination scenario. 
In order to distinguish the two cases, a precise knowledge of the spatial origin of the VHE emission is essential.
X-ray emission has been detected from the star
forming regions W51A and W51B. W51C is also visible in X-rays
showing both a shell type and center-filled morphology~\cite{Koo2002}. A non-thermal
X-ray emission has only been detected from the relatively bright and compact
source CXO J192318.5+140505, which is thought to be a pulsar wind
nebula (PWN) associated to the SNR~\cite{Koo2005}.

{\em Fermi}/LAT detected gamma-ray emission from 200MeV to 50GeV
extended throughout W51B and W51C~\cite{FermiW51}. The relatively
large point-spread-function PSF does not allow to tell which objects of the
field of view is responsible for the observed emission. 
The {\em Fermi}/LAT emission
region is extended in comparison to the PSF of the
instrument ($\sigma_{\mathrm{ext}} = 0.22 \pm 0.02 \,\mathrm{deg}$. The H.E.S.S. Collaboration detected VHE emission above 420~GeV from
an extended region coincident with the {\em Fermi}/LAT emission
region~\cite{HESS-W51}. In their sky map, the H.E.S.S. emission is
smoothed with a radius of 0.22$^\circ$~\cite{HESS-W51} and, as well as
in the case of {\em Fermi}/LAT, it overlaps with several HII regions,
the molecular mass in W51B as well as the PWN candidate
CXOJ192318.5+140305. The flux measured by H.E.S.S. above 1~TeV is at
the level of 3\% of the Crab Nebula flux. Finally, also the MILAGRO
Collaboration reported a possible excess from the same source at
energies above several TeV~\cite{Milagro-W51}.
The modeling of the spectral energy distribution (SED) measured by
{\em Fermi}/LAT suggests a hadronic mechanism as the main origin
of the gamma rays. The interaction should take place in the region
between the supernova remnant and the dense, complex region North-West
of it. In this case, the engulfed cloud would be those described in~\cite{Koo1997a}\cite{Koo1997b}.

\section{Observations with the MAGIC telescopes}

MAGIC consists of two 17~m diameter imaging atmospheric Cherenkov
telescopes  located at the Roque de los Muchachos in the Canary Island
of La Palma ($28^{\circ}46'$N, $17^{\circ}53'$W) at the height of
2200~m\,a.s.l. Astronomical observations of VHE gamma-ray sources are
performed with the two telescopes simultaneously which provides a major
improvement of performance with respect to the single telescope
observations previously done with MAGIC-I~\cite{StereoPaper}.
MAGIC observed W51C between May 17 and August 19 2010. The
observations were carried out in the so-called wobble mode and covered
a zenith angle range between 14 and 35 degrees. The {\em Fermi}/LAT source W51C (RA=19.385~h,
$\delta$=14.19~$^\circ$) was chosen as central position
for the observations. After applying quality cuts we
collected a total of 31.1~h effective\footnote{Due to dead time of the 
MAGIC-II telescope readout system the effective observation time is 
always lower than the real observation time.} dark time.  All data
were taken in stereoscopic mode, recording only events that triggered both
telescopes. 
The trigger energy threshold of the system is around
50~GeV~\cite{StereoPaper}. This is the lowest energy threshold among
IACTs in operation and provides the chance to have ground based
observations with an energy range overlapping with the one of {\em Fermi}/LAT.
The analysis of the data was performed using the MARS analysis framework which
is the standard software used for MAGIC data
analysis~\cite{MoralejoLodz}. The details of the analysis, as well as
the general performance of MAGIC in stereoscopic mode, are reported
in~\cite{StereoPaper}~\cite{StereoICRC}.

\section{Results from MAGIC observations}

\begin{figure}[!t]
  \vspace{5mm} \centering
  \includegraphics[width=0.95\columnwidth]{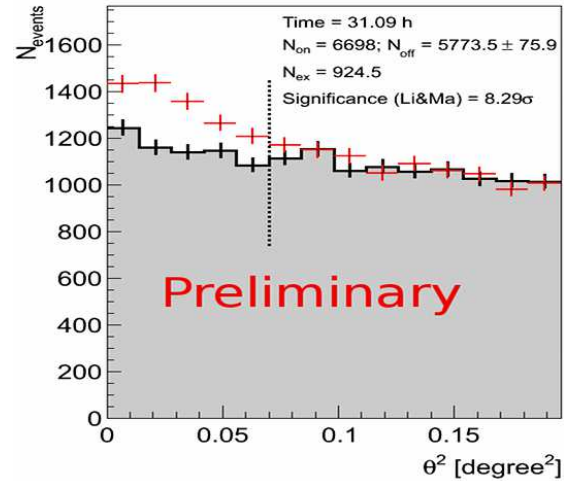}
  \caption{Distribution of the squared angular distance between the
    reconstructed VHE gamma-like events (red
    points) above 150~GeV and
    the source position together with background (black points) reconstructed from a 
symmetric position in the camera(s). The vertical dotted line marks the
  region defined to compute the excess events. }
  \label{Theta2}
 \end{figure}

\begin{figure}[!t]
  \vspace{5mm}
  \centering
  \includegraphics[width=0.95\columnwidth]{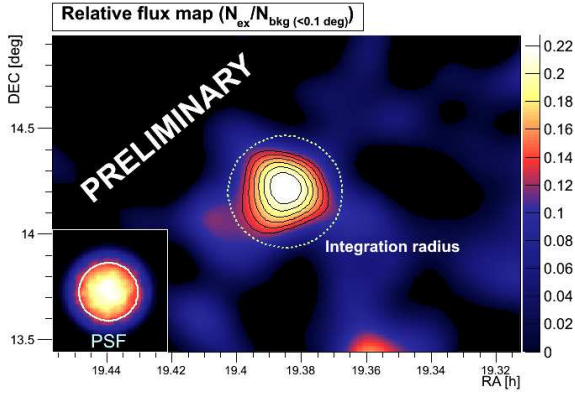}
  \caption{VHE gamma-ray emission from W51C obtained with the MAGIC
    telescopes above 150 GeV. The map has been smoothed with a
    Gaussian kernel of $\sigma$=0.10$^\circ$. The color scale shows
    the relative flux of gamma rays (excess events normalized to the
    number of background events) and the black contours are different
    levels of the test statistic variable (Li \& Ma
    eq. 17~\cite{LiMa83} applied on a smoothed and modeled background
    estimation. It roughly corresponds to a Gaussian significance.
    More details given in~\cite{Saverio2011}). In the figure the region
    defined for integrating the signal (dotted line) and the PSF after
    smearing (bottom left inset) are also shown.}
  \label{SkymapMagic}
 \end{figure}

MAGIC founds a spatially extended excess of VHE gamma-ray events
from the direction of W51C. In total 924 excess events were found above
150~GeV in the 31.1~hours of 
effective time (see figure~\ref{Theta2}), resulting an a detection significance of 8.29~$\sigma$. 
The source as seen by MAGIC shows an extension (sigma of two-dimensional Gaussian fit) of
0.16$\pm$0.02$^\circ$. This lies well above the PSF of
the analysis (0.08$^\circ$ in the same energy range assuming a spectral
index of -2.6). The centroid of the emission detected by MAGIC is
coincident with the centroid of the {\em Fermi}/LAT emission:
RA=19.387$\pm$0.002~h, $\delta$=14.18$\pm$0.02$^\circ$. A sky map of
the MAGIC observation at energies above 150~GeV is shown in
figure~\ref{SkymapMagic}. Details about MAGIC sky maps and test
statistic are given in~\cite{Saverio2011}.

\begin{figure}[!t]
  \vspace{5mm}
  \centering
  \includegraphics[width=0.95\columnwidth]{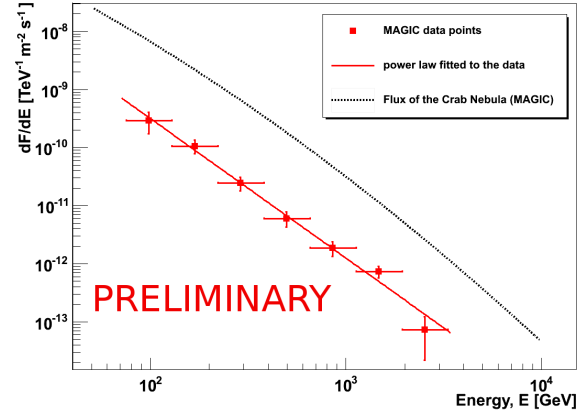}
  \caption{Measured flux from W51C with the MAGIC telescopes. The red
    points show the differential fluxes. The red line represents the
    best fitted power-law. As a reference the measured by MAGIC-I
    differential energy spectrum from the Crab Nebula is also shown.}
  \label{Spectrum}
 \end{figure}

MAGIC has measured the differential energy spectrum of the VHE
gamma-ray emission in the energy range 75~GeV to 3.3~TeV. The
differential spectrum is well fitted by a power law as can be
seen in figure~\ref{Spectrum} ($\chi^2/ndf$ = 4.5/5). The
obtained spectral index of the  VHE gamma-ray emission is
-2.40$\pm$0.12$_{stat}$. The integrated flux above 1 TeV corresponds to 3.8\% of the
Crab Nebula in agreement with the integral flux reported by
H.E.S.S.~\cite{HESS-W51} above
1 TeV. The obtained flux in units of TeV$^{-1}$ cm$^{-2}$ s$^{-1}$
is given by: 

\begin{eqnarray*}
\frac{dF}{dE} = \mathrm {(1.25 \pm 0.18_{ stat}) \times 10^{-12}}
\left ( \frac{E}{TeV}\right ) ^ \mathrm{(-2.40\pm0.12_{stat})}
\end{eqnarray*}

\section{Discussion}

\begin{figure}[!t]
  \vspace{5mm}
  \centering
  \includegraphics[width=0.95\columnwidth]{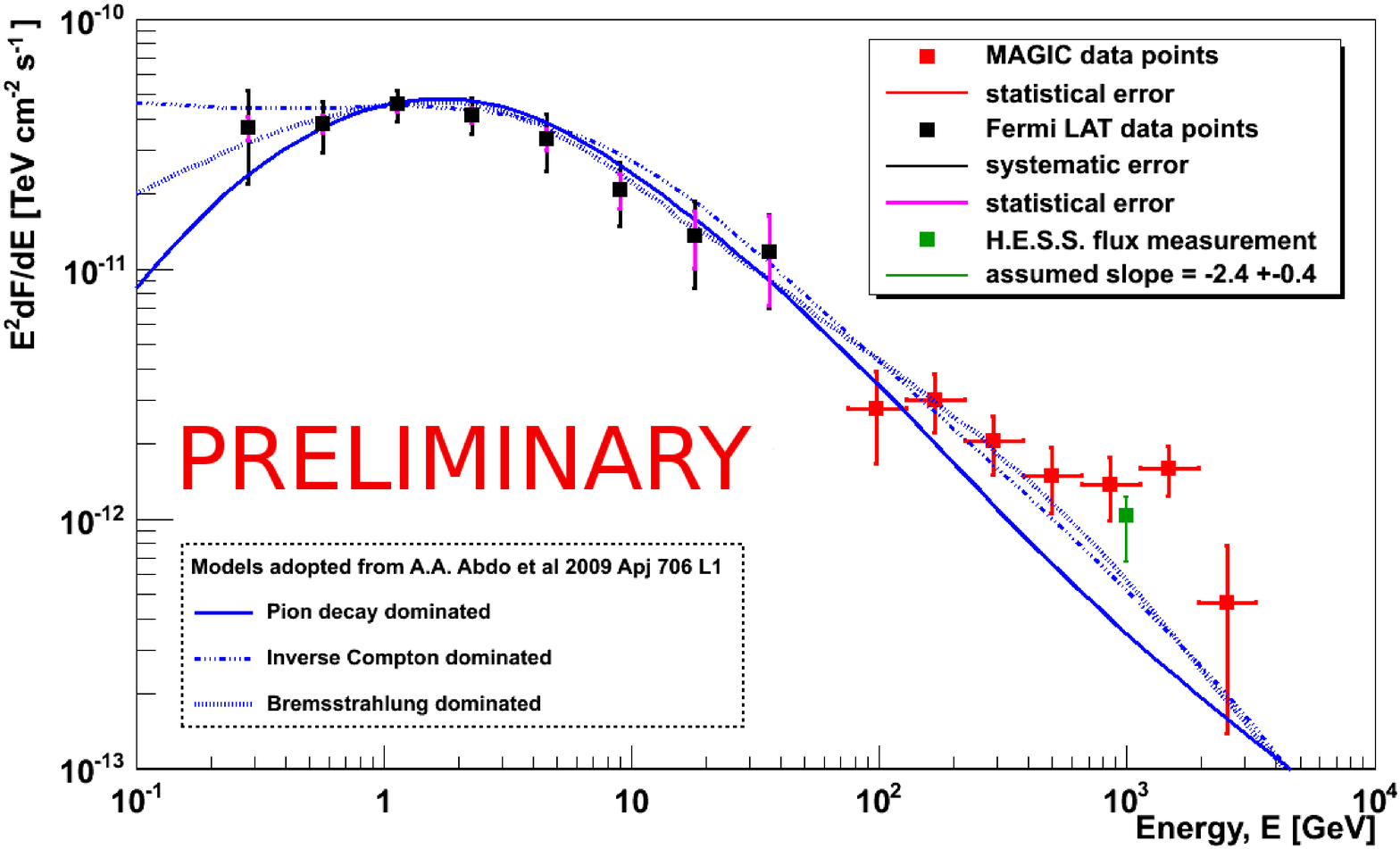}
  \caption{SED of W51C measured by {\em Fermi}/LAT (black points), MAGIC (red
    points) and H.E.S.S. (green point). Also shown in the figure the 3
  different scenarios for modeling the multi wavelength data that were
  used in~\cite{FermiW51}. Gamma-ray emission is explained by
  emission coming from a hadronic dominated scenario (continuous blue
  line), inverse Compton dominated scenario (dot-dashed line) or a
  Bremsstrahlung dominated scenario (dashed line). }
  \label{SED}
 \end{figure}

The MAGIC data do not only fill the gap between the {\em Fermi}/LAT and the
H.E.S.S. measurements but furthermore show that the spectral distribution can be described by a single power-law
from energies of $\sim$10\,GeV up to TeV energies. Figure~\ref{SED} shows the SED measured by
{\em Fermi}/LAT and MAGIC together with the H.E.S.S. measurement converted into
a differential flux\footnote{The integral flux above 1 TeV reported 
in~\cite{HESS-W51} is converted into a differential flux using 
the MAGIC spectral index of -2.4. An error of $\pm0.4$ is assumed in 
order to obtain the error bars shown.}. MAGIC data
agrees well with the {\em Fermi}/LAT and H.E.S.S. measurements. In the same
figure the predictions from the phenomenological model used by {\em Fermi}/LAT
in~\cite{FermiW51} to explain the origin of the gamma-ray emission are
also shown. Three different scenarios are considered: one where
gamma-ray emission is dominated by $\pi^0$ decay and two more where
the gamma-ray emission is mainly dominated by inverse Compton (IC) or
Bremsstrahlung of electrons.

The VHE MAGIC data points as well as H.E.S.S. point are
above the predictions from the model in all scenarios at energies above 1
TeV. Although this seems to favor the electromagnetic scenarios, both leptonic models show 
difficulties in fitting the radio data and, in the case of the IC model, severe energy constraints.
In addition both leptonic models need an electron to proton ratio in order of one. Such a value is much higher than the one observed in direct cosmic ray measurements 
(e/p $\sim 10^{-3}$) and difficult to explain due to the much higher energy losses for electrons.
Concerning the hadronic emission the parameters used by the \textit{Fermi}/LAT-team leads to a 
proton spectral index between $\sim$10 and $\sim$500\,GeV close to -3.  
A slightly harder index would be able to explain both MAGIC and the \textit{Fermi}/LAT points.
The spectral index measured by \textit{Fermi}/LAT above 10 GeV is $-2.5\pm0.18$ \citep{david}.
Taken this into account the non matching of this hadronic scenario does not rule out a hadronic origin of the emission. 
Moreover a hadronic origin predicts a constant slope of the gamma emission above several GeV up to the cut-off originating from reaching the maximum energy
of protons responsible for the emission.

\begin{figure*}[htb]
  \centering
  \includegraphics[width=6in,height=3in]{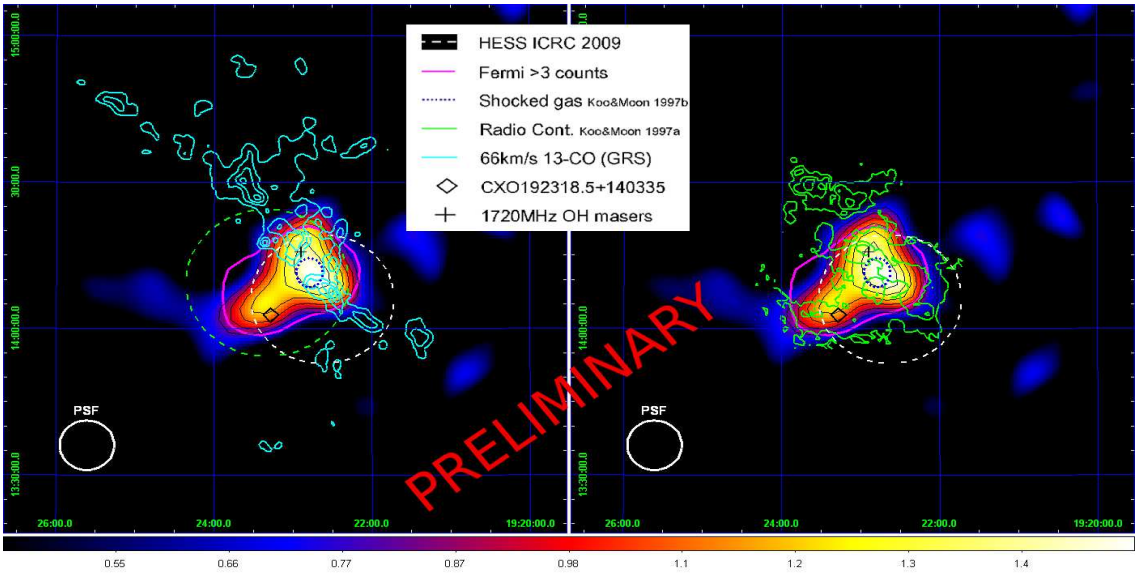}
  \caption{Map of W51C region in different wavelengths. Shown in
    colors the MAGIC relative flux above $700\,\mathrm{GeV}$ (smooth by a Gaussian kernel of
    0.065$^\circ$) overlapped with the test statistics significance
    contours in black. The pink line shows the approximate contour of
    the HE emission detected by {\em Fermi}/LAT and the white dashed line shows
    the approximate contour of the H.E.S.S. VHE emission. The green
    dashed line shows the approximate contour of the SNR W51C. The
    dotted dark blue line shows the shocked gas region defined by Koo
    et al.~\cite{Koo1997b}. On the left map the molecular clouds measured with the
    66~km/s 13-CO line are shown in light blue. On the right map, the
    green contours show the radio data from Koo \&
    Moon~\cite{Koo1997a} in green. The X-ray source and PWN candidate
    CXO192318.5+140335 and a 1720MHz OH maser are also shown in the
    map.}
  \label{MW_Map}
 \end{figure*}

The angular resolution of MAGIC at energies of 100~GeV is comparable
to that of {\em Fermi}/LAT. For energies above 700~GeV, the
angular resolution MAGIC is as low as
$\simeq0.05^\circ$. This allows for a unprecedented resolution sky map from
W51C. Figure~\ref{MW_Map} shows the MAGIC view of W51C overlapped with
data from different wavelengths.
The overall MAGIC emission above 700~GeV
is contained within the region of 3 counts above 1\,GeV by {\em Fermi}/LAT (pink line
in the figure). The higher angular resolution resolves that the gross of the VHE gamma-ray
emission above 700~GeV is spatially coincident with the shocked gas region
reported by Koo et al. in~\cite{Koo1997b}.
This also supports the hadronic origin of the HE-VHE gamma
rays. Furthermore there is an extension roughly following the shape of the remnants shell in radio and the {\em Fermi}/LAT HE
emission. In addition this VHE emission towards the South-East goes in the direction of the PWN
candidate.\\ 

\section{Conclusions}
MAGIC observations do not only confirm the emission of HE-VHE gamma rays from an
extended source located in the SNR W51C, but extend the spectrum almost without gap above the 
\textit{Fermi}/LAT energies. The measured SED is in agreement with a extrapolation of the Fermi data and agrees
with the flux measured by the H.E.S.S. collaboration.  
The emission measured by MAGIC is spatially coincident with that reported by {\em Fermi}/LAT. 
Moreover, the higher angular resolution
provided by MAGIC shows that the bulk of the VHE gamma-ray emission
comes from the shocked molecular cloud located where the SNR shock
engulfs a large molecular cloud, creating a shocked gas region
distinguishable in the radio data. This fact, a constant spectral index of the emission from $\sim$GeV to multi-TeV, and 
the problems to explain the emission in a leptonic scenario suggest that the gamma-ray
emission has most likely a hadronic origin as it is expected in the
case of gamma rays produced in a {\em crushed cloud} scenario.

\begin{acknowledgements}
We would like to thank the Instituto de Astrof\'{\i}sica de
Canarias for the excellent working conditions at the
Observatorio del Roque de los Muchachos in La Palma.
The support of the German BMBF and MPG, the Italian INFN, 
the Swiss National Fund SNF, and the Spanish MICINN is 
gratefully acknowledged. This work was also supported by 
the Marie Curie program, by the CPAN CSD2007-00042 and MultiDark
CSD2009-00064 projects of the Spanish Consolider-Ingenio 2010
programme, by grant DO02-353 of the Bulgarian NSF, by grant 127740 of 
the Academy of Finland, by the YIP of the Helmholtz Gemeinschaft, 
by the DFG Cluster of Excellence ``Origin and Structure of the 
Universe'', by the DFG Collaborative Research Centers SFB823/C4 and SFB876/C3,
and by the Polish MNiSzW grant 745/N-HESS-MAGIC/2010/0.
\end{acknowledgements}



\newpage

\begin{thebibliography}{}
\bibitem{david} Paneque D. et al., Fermi Symposium, Sources in the Fermi Sky above 10 GeV, 2011
\bibitem{Reynolds2008} Reynolds S.P., ARA\&A 46, 89–126, 2008
\bibitem{Hinton2010}  Hinton J. 2009. ARA\&A 47, 523-565, 2009
\bibitem{HESS-W28} Aharonian F et al., A\&A 481, 401–410, 2008
\bibitem{MAGIC-IC443} Albert J et al., ApJ 664, 87–90, 2007
\bibitem{Uchiyama2010} Uchiyama Y. et al, ApJL, 723, 122-126, 2010
\bibitem{Koo2002} Koo B.-C., et al., AJ, 123, 1629-1638, 2002
\bibitem{Koo2005} Koo, B.-C., Lee, J.-J., Seward, F. D., \& Moon, D.-S, ApJ, 633, 946, 2005
\bibitem{Koo1995} Koo B.-C., Kim K.-T., Seward, F. D., ApJ, 447, 211, 1995
\bibitem{Koo1997a} Koo B.-C., \& Moon, D.-S., ApJ, 475, 194, 1997
\bibitem{Koo1997b} Koo B.-C., \& Moon, D.-S., ApJ, 485, 263, 1997
\bibitem{FermiW51} Abdo A.A. et al., ApJ 706, 2009, L1–L6, 2009
\bibitem{HESS-W51} Feinstein F. et al., AIP Conf. Proc., 1112, 54-62, 2009
\bibitem{Milagro-W51} Abdo, A. A., et al., ApJ, 700, L127, 2009
\bibitem{StereoICRC} E. Carmona, J. Sitarek, P. Colin, M. Doert,
  S. Klepser, S. Lombardi, M. L\'opez, A. Moralejo, S. Pardo,
  V. Scalzotto, R. Zanin et al.,  Proceedings of the 32nd International Cosmic Ray Conference, 2011, arXiv:1110.0947v1
\bibitem{StereoPaper} Aleksi\'c, J. et al. (MAGIC Collaboration), submitted to A\&A, 2011, arXiv:1108.1477v1
\bibitem{MoralejoLodz} Moralejo A. et al., Proceedings of 31st
  International Cosmic Ray Conference, 2009, arXiv:0907.0943v1 
\bibitem{Saverio2011}  S. Lombardi, K. Berger, P. Colin, A.-D. Ortega,
  S. Klepser et al., Proceedings of the 32nd International Cosmic Ray Conference, 2011, arXiv:1109.6195v1
\bibitem{LiMa83} Li, T.-P. \& Ma, Y.-Q., ApJ, 272, 317, 1983


\end{thebibliography}
\end{document}